\documentclass[twocolumn,showpacs,preprintnumbers,floatfix,
               amsmath,amssymb]{revtex4}
\usepackage{longtable}
\usepackage{graphics}
\usepackage{epsfig}
\def\pmbf#1{{\mbox{\boldmath${#1}$}}}
\newcommand{\rvec}{\mathbf{r}}

\begin{document}

\title{Nonuniform Neutron-Rich Matter and
       Coherent Neutrino Scattering}

\author{C.J. Horowitz}\email{horowit@indiana.edu} 
\author{M.A. P\'{e}rez-Garc\'{\i }a}\email{mperezga@indiana.edu}
\author{J. Carriere}
\affiliation{Nuclear Theory Center and Department of Physics, 
             Indiana University,  Bloomington, IN 47405}
\author{D. K. Berry}
\affiliation{University Information Technology Services,
             Indiana University, Bloomington, IN 47408}
\email{dkberry@indiana.edu}
\author{J. Piekarewicz}\email{jorgep@csit.fsu.edu}
\affiliation{Department of Physics,
         Florida State University, Tallahassee, FL 32306}


\date{\today}

\begin{abstract}
Nonuniform neutron-rich matter present in both core-collapse
supernovae and neutron-star crusts is described in terms of a
semiclassical model that reproduces nuclear-matter properties and
includes long-range Coulomb interactions. The neutron-neutron
correlation function and the corresponding static structure factor are
calculated from molecular dynamics simulations involving 40,000 to
100,000 nucleons. The static structure factor describes coherent
neutrino scattering which is expected to dominate the neutrino
opacity. At low momentum transfers the static structure factor is
found to be small because of ion screening. In contrast, at
intermediate momentum transfers the static structure factor displays a
large peak due to coherent scattering from all the neutrons in a
cluster. This peak moves to higher momentum transfers and decreases in
amplitude as the density increases.  A large static structure factor
at zero momentum transfer, indicative of large density fluctuations
during a first-order phase transition, may increase the neutrino
opacity. However, no evidence of such an increase has been found.
Therefore, it is unlikely that the system undergoes a simple
first-order phase transition. Further, to compare our results to more
conventional approaches, a cluster algorithm is introduced to
determine the composition of the clusters in our simulations.
Neutrino opacities are then calculated within a single heavy nucleus
approximation as is done in most current supernova simulations.  It is
found that corrections to the single heavy nucleus approximation first
appear at a density of the order of $10^{13}$ g/cm$^3$ and increase
rapidly with increasing density. Thus, neutrino opacities are
overestimated in the single heavy nucleus approximation relative to
the complete molecular dynamics simulations.
\end{abstract}
\maketitle
%
\section{Introduction}
\label{intro}

The description of nuclear matter at subnuclear densities is an
important and general problem.  Attractive short-range strong
interactions correlate nucleons into nuclei. However, nuclear sizes
are limited by long-range repulsive Coulomb interactions and thermal
excitations.  This competition between attraction and repulsion
produces multifragmentation in heavy ion collisions; the breaking 
of the system into intermediate sized 
fragments~\cite{Bod80_PRC22,Sch87_PRC36,Pei91_PLB260}.
In astrophysics this competition produces a variety of complex
phenomena.  At densities considerably lower than normal nuclear
matter saturation density, the system may be described as a collection of
nearly free nucleons and nuclei in nuclear statistical equilibrium
(NSE), while at normal nuclear matter saturation density and above the
system is expected to become uniform. In between these regimes {\it
pasta phases} may develop with nucleons clustered into subtle and
complex shapes~\cite{Rav83_PRL50,Has84_PTP71}. This pasta phase may 
be present in the inner crust of neutron stars and in core collapse
supernovae.  Unfortunately, low-density NSE models and high-density
models of nuclear matter are often incompatible.  As the density
increases, it is difficult to account for the strong interactions
between nuclei in NSE models.  Likewise, as the density decreases and
the uniform system becomes unstable against fragmentation, uniform
models fail to describe cluster formation.

We wish to study the different phases and properties of the system as
a function of density. This is essential in the simulations of
core-collapse supernovae as they involve a tremendous range of
densities and temperatures.  Several semiclassical simulation
techniques, often developed for heavy ion collisions, may be used to
describe the system over this large density range. Indeed, Watanabe
and collaborators have used quantum molecular dynamics to describe the
structure of the pasta~\cite{Wat03_PRC68}. These simulations should
reduce to isolated nuclei at low densities and to uniform matter at
high densities. In principle a first order phase transition could have
a two phase coexistence region.  Large density fluctuations at the
transition could greatly increase the neutrino
opacity~\cite{Mar04_xxx}.  In the present paper, we search for regions
with large density fluctuations using molecular dynamics simulations.

The main focus of the present paper is coherent neutrino scattering,
an essential tool for computing neutrino mean-free paths in supernovae
and to determine how neutrinos are initially trapped.  Furthermore,
coherent neutrino scattering---by representing long-range
``classical'' physics---may provide insights into how the clustering
evolves with density in a model-independent way. In a previous paper a
simple Monte-Carlo simulation model involving 4,000 particles was
developed and first results for neutrino mean-free paths were
presented~\cite{Hor04_PRC69}. In this paper results are presented for
larger simulations involving up to 100,000 nucleons using molecular
dynamics. This large number of nucleons is required to accommodate long
wavelength neutrinos. For example, the wavelength of a 10 MeV neutrino
is approximately 120 fm.  At a baryon density of 0.05 fm$^{-3}$, a
simulation volume of one neutrino wavelength per side contains close
to 100,000 nucleons.
                        
Having generated a variety of observables, our microscopic
results are then compared to those generated from a macroscopic
cluster model. Such macroscopic models describe the system as a
collection of free nucleons plus a single species of a heavy nucleus
and are presently used in most supernovae simulations.  By comparing
the two approaches, we gain insight into the strengths and limitations
of the macroscopic cluster models. There is a duality between
microscopic descriptions of the system in terms of nucleon coordinates
and macroscopic descriptions in terms of effective nuclear degrees of
freedom. Thus, it is interesting to learn when does a neutrino scatter
coherently from a nucleus and when does it scatter from an individual
nucleon? At the Jefferson Laboratory a similar question is being
posed: when does a photon couple coherently to a full hadron and when
to an individual quark? The quark/hadron duality has provided insight
on how descriptions in terms of hadron degrees of freedom can be
equivalent to descriptions in terms of quark
coordinates~\cite{Jes02_PRD65}. Here we are interested in
nucleon/nuclear duality: how can nuclear models incorporate
the main features of microscopic nucleon descriptions?

The manuscript has been organized as follows. In Sec.~\ref{formalism}
the simple semiclassical formalism is introduced as well as details of
the molecular dynamics simulations. In Sec.~\ref{scattering} we review
the formalism for neutrino scattering and relate it to the static
structure factor.  Simulation results are presented in
Sec.~\ref{results} including the calculation of neutrino mean-free
paths using nucleon coordinates.  Section~\ref{cluster_model} presents
a simple cluster model that compares these results to more
conventional approaches using {\it nuclear} coordinates.  Finally,
conclusions and future directions are presented in
Sec.~\ref{conclusions}.

\section{Formalism}
\label{formalism}

In this section we review our semiclassical model that while simple,
contains the essential physics of competing interactions consisting of
a short-range nuclear attraction and a long-range Coulomb repulsion.
The impossibility to simultaneously minimize all elementary
interactions is known in condensed-matter circles as {\it
frustration}. The complex physics of frustration, along with many
other details of the model, may be found in Ref.~\cite{Hor04_PRC69}.  Here
only a brief review of the most essential features of the model is
presented. We model a charge-neutral system of electrons, protons, and
neutrons. The electrons are assumed to be noninteracting and thus are
described as a degenerate free Fermi gas at a number density identical
to that of the protons ({\it i.e.,} $\rho_e\!=\!\rho_p$).  The
nucleons, on the other hand, interact classically via a
nuclear-plus-Coulomb potential. However, the use of an effective
temperature and effective interactions are used to simulate effects
associated with quantum zero-point motion. More elaborate models are
currently under construction and these will be presented in future
contributions.  While simple, the model displays the essential physics
of frustration, namely, nucleons clustering into pasta but the size of 
the clusters limited by the Coulomb repulsion, in a transparent
form. Moreover, one may study the evolution of the system through the
low density, pasta, and high density phases within a single
microscopic model. Finally, the model facilitates simulations with a
large numbers of particles, a feature that is essential to estimate
and control finite-size effects and, as alluded earlier, to reliably
study the response of the system to long wavelength neutrinos.

The total potential $V_{\rm tot}$ energy of the system consists of a
sum of two-body interactions
\begin{equation}
 V_{tot}=\sum_{i<j} V(i,j) \;,
\label{vtot}
\end{equation}
where the ``elementary'' two-body interaction is given as follows:
\begin{equation}
V(i,j) = a e^{-r_{ij}^{2}/\Lambda} + \Big[b+c\tau_z(i)\tau_z(j)\Big]
       e^{-r_{ij}^{2}/2\Lambda}+V_{\rm c}(i,j)\;.
\label{v}
\end{equation}
Here the distance between the particles is denoted by $r_{ij}=|{\bf
r}_i\!-\!{\bf r}_j|$ and $\tau_z$ represents the nucleon isospin
projection ($\tau_z\!=\!+\!1$ for protons and $\tau_z\!=\!-\!1$ for
neutrons).  The two-body interaction contains the characteristic
intermediate-range attraction and short-range repulsion of the
nucleon-nucleon force. Further, an isospin dependence has been
incorporated in the potential to ensure that while pure neutron matter
is unbound, symmetric nuclear matter is appropriately bound. Indeed,
the four model parameters ($a$, $b$, $c$, and $\Lambda$) introduced in
Eq.~({\ref v}) have been adjusted in Ref.~\cite{Hor04_PRC69} to reproduce
the following bulk properties: a) the saturation density and binding
energy per nucleon of symmetric nuclear matter, b) (a reasonable value
for) the binding energy per nucleon of neutron matter at saturation
density, and c) (approximate values for the) binding energy of a few
selected finite nuclei. All these properties were computed via a
classical Monte Carlo simulation with the temperature arbitrarily
fixed at 1 MeV. The parameter set employed in all previous and present
calculations is displayed in Table~\ref{Table1}.
Finally---and critical for pasta formation---a screened Coulomb
interaction of the following form is included:
\begin{equation}  
V_{\rm c}(i,j)=\frac{e^{2}}{r_{ij}}e^{-r_{ij}/\lambda}
                 \tau_p(i)\tau_p(j) \;,
\label{vc}
\end{equation}
where $\tau_p\!\equiv\!(1\!+\!\tau_z)/2$ and $\lambda$ is the
screening length that results from the slight polarization of the
electron gas. The relativistic Thomas-Fermi screening length is given
by
\begin{equation}
 \lambda=\frac{\pi}{e} 
 \left(k_{\rm F}\sqrt{k_{\rm F}^2+m_e^2}\right)^{-1/2}
 \hspace{-0.2cm}\;,
 \label{lambda}
\end{equation}
where $m_e$ is the electron mass and the electron Fermi momentum has
been defined by $k_{\rm
F}\!=\!(3\pi^2\rho_e)^{1/3}$~\cite{Fet71_MH,Chi77_AOP108}. Unfortunately,
while the screening length $\lambda$ defined above is smaller than the
length $L$ of our simulation box, it is not significantly
smaller. Hence, following a prescription introduced in
Ref.~\cite{Hor04_PRC69} in an effort to control finite-size effects, the
value of the screening length is arbitrarily decreased to
$\lambda\!=\!10$~fm.

The simulations are carried out with both a fixed number of particles
$A$ and a fixed density $\rho$. The simulation volume is then simply
given by $V\!=\!A/\rho$. To minimize finite-size effects periodic
boundary conditions are used, so that the distance $r_{ij}$ is
calculated from the $x$, $y$, and $z$ coordinates of the $i_{\rm th}$
and $j_{\rm th}$ particles as follows:
\begin{equation}
 r_{ij}=\sqrt{[x_i-x_j]^2 + [y_i-y_j]^2 + [z_i-z_j]^2} \;,
 \label{pbc}
\end{equation}
where the periodic distance, for a cubic box of side $L=V^{1/3}$,
is given by 
\begin{equation}
[l]={\rm Min}( |l|, L-|l| ) \;.
\end{equation}

In our earlier work~\cite{Hor04_PRC69} properties of the pasta were
obtained from a partition function that was calculated using
Monte Carlo integration.  In the present paper molecular dynamics is
used to simulate the system.  There are a few advantages in using
molecular dynamics over a partition function. First, larger systems
are allowed to be simulated due to advances in both software and in
hardware (see appendix). Second, one is not limited to compute the
static structure factor of the system as now the full dynamic response
is available. One expects the neutron-rich pasta to display
interesting low-energy collective excitations---such as {\it Pygmy
resonances}---that may be efficiently excited by low-energy
neutrinos. These low-energy modes of the pasta are currently under
investigation. 

To carry out molecular dynamics simulations the
trajectories of all of the particles in the system are determined by
simply integrating Newton's laws of motion, albeit for a very large
number of particles (up to 100,000 in the present case) using the
velocity-Verlet
algorithm~\cite{Erc97_WWW,Ves01_Kluwer,All03_Oxford}. To start the
simulations, initial positions and velocities must be specified for
all the particles in the system. The initial positions are randomly
and uniformly distributed throughout the simulation volume while the
initial velocities are distributed according to a Boltzmann
distribution at temperature $T$.  As the velocity-Verlet is an
energy---not temperature---conserving algorithm, kinetic and potential
energy continuously transformed into each other.  To prevent these
temperature fluctuations, the velocities of all the particles are
periodically rescaled to ensure that the average kinetic energy per 
particle remains fixed $(3/2)k_{\rm B}T$.

In summary, a classical system has been constructed with a total
potential energy given as a sum of two-body, momentum-independent
interactions as indicated in Eq.~(\ref{v}). Expectation values of any
observable of interest may be calculated as a suitable time average
using particle trajectories generated from molecular dynamics
simulations.

\begin{table}
\caption{Model parameters used in the calculations.}
 \begin{ruledtabular}
 \begin{tabular}{cccc}
   $a$     & $b$     & $c$    & $\Lambda$ \\
   \hline
   110 MeV & -26 MeV & 24 MeV & 1.25 fm$^2$
 \label{Table1}
 \end{tabular}
\end{ruledtabular}
\end{table}

\section{Neutrino Scattering}
\label{scattering}

In this section we review coherent neutrino scattering which is
expected to dominate the neutrino opacity in regions where clusters,
such as either nuclei or pasta, are present. Although the formalism
has been presented already in Ref.~\cite{Hor04_PRC69}, some details are
repeated here (almost verbatim) for the sake of completeness and
consistency.

In the absence of corrections of order $E_{\nu}/M$ (with $E_\nu$ the
neutrino energy and $M$ the nucleon mass) and neglecting contributions
from weak magnetism, the cross section for neutrino-nucleon elastic
scattering in free space is given by the following
expression~\cite{Hor02_PRD65}:
\begin{equation}
 \frac{d\sigma}{d\Omega} = \frac{G_{F}^2 E_\nu^2}{4\pi^2} 
 \left[c_a^2 (3-\cos\theta) + 
       c_v^2 (1+\cos\theta)\right] \;,
 \label{cross}
\end{equation}
where $G_{F}$ is the Fermi coupling constant and $\theta$ the
scattering angle.

Having neglected the contribution from weak magnetism, the weak
neutral current $J_\mu$ of a nucleon contains only axial-vector
($\gamma_5\gamma_\mu$) and vector $\gamma_\mu$ contributions. That is,
\begin{equation}
  J_\mu=c_a\gamma_5\gamma_\mu + c_v\gamma_\mu \;.
\end{equation}
The axial coupling constant is,
\begin{equation}
 c_a=\pm \frac{g_a}{2} \quad (g_a=1.26)\;.
\label{ca}
\end{equation}
Note that in the above equation the $+(-)$ sign is for
neutrino-proton(neutrino-neutron) scattering. The weak charge of the
proton $c_{v}$ is small, as it is strongly suppressed by the
weak-mixing (or Weinberg) angle $\sin^{2}\theta_{\rm W}\!=\!0.231$. It
is given by  
\begin{equation}
 c_v=\frac{1}{2}-2\sin^{2}\theta_{\rm W}=0.038 \approx 0\;.
\label{cvp}
\end{equation}
In contrast, the weak charge of a neutron is both large and
insensitive to the weak-mixing angle: $c_v\!=\!-\!1/2$.

If nucleons cluster tightly, either into nuclei or into pasta, then
the scattering of neutrinos from the various nucleons in the cluster
may be coherent. As a result, the cross section will be significantly
enhanced as it would scale with the {\it square} of the number of
nucleons~\cite{Fre77_ARNS27}. In reality, only the contribution from
the vector current is expected to be coherent. This is due to the 
strong spin and isospin dependence of the axial current, which is 
expected reduce its coherence. (Recall that in the nonrelativistic 
limit, the nucleon axial-vector current becomes 
$\gamma_5{\pmbf\gamma}\tau_{z}\rightarrow {-\pmbf\sigma}\tau_{z}$). 
Since in nuclei and presumably also in the pasta most nucleons pair 
off into spin singlet states, their axial-vector coupling to neutrinos
will be strongly reduced. Hence, in this work we focus
exclusively on coherence effects from the vector current. Coherence is
important in neutrino scattering from the pasta because the neutrino
wavelength is comparable to the interparticle spacing and even to the
intercluster separation. One must then calculate the relative phase
for neutrino scattering from different nucleons and then add their
contribution coherently. This procedure is embodied in the static
structure factor $S(q)$.

The static structure factor {\it per neutron} is defined as follows:
\begin{equation}
  S({\bf q})=\frac{1}{N}
             \sum_{n\neq 0}\Big|\langle\Psi_{n}|\hat{\rho}
             ({\bf q})|\Psi_{0}\rangle\Big|^{2} \;,
\label{sq}
\end{equation}
where $\Psi_{0}$ and $\Psi_{n}$ are ground and excited nuclear
states, respectively and the weak vector charge density is given 
by
\begin{equation}
 {\rho}({\bf q})=\sum_{i=1}^{N} 
  \exp(i{\bf q}\cdot{\bf r}_i) \;.
 \label{rhoq}
\end{equation}
As the small weak charge of the proton [Eq.~(\ref{cvp})] will be
neglected henceforth, the sum in Eq. (\ref{rhoq}) runs only over the
$N$ neutrons in the system. The cross section {\it per neutron} for
neutrino scattering from the whole system is now given by
\begin{equation}
 \frac{1}{N}\frac{d\sigma}{d\Omega}=S({\bf q}) 
 \frac{G_{F}^2 E_\nu^2}{4\pi^2}\frac{1}{4}(1+\cos\theta)\;.
 \label{sigma2}
\end{equation}
The above expression is the single neutrino-neutron cross section per
neutron obtained from Eq.~(\ref{cross}) (with $c_a\!\equiv \!0$)
multiplied by S({\bf q}). This indicates that S({\bf q}) contains the
effects from coherence. Finally, note that the momentum transfer is
related to the scattering angle through the following equation:
\begin{equation}
 q^2=2E_\nu^2(1-\cos\theta) \;.
\label{q}
\end{equation}
The static structure factor has important limits. For a detailed
justification of these limits the reader is referred to
Ref.~\cite{Hor04_PRC69}. In the limit that the momentum transfer to the
system goes to zero ($q\!\rightarrow\!0$) the weak charge density
[Eq.~(\ref{rhoq})] becomes the neutron number operator
$\hat{\rho}(q\!=\!0)\!=\!\hat{N}$. In this limit the static structure
factor reduces to,
\begin{equation}
 S(q=0)=\frac{1}{N}\left(\langle\hat{N}^{2}\rangle-
                         \langle\hat{N}\rangle^{2}\right).
 \label{S0}
\end{equation}
Thus, the $q\!\rightarrow\!0$ limit of the static structure factor is
related to the fluctuations in the number of particles, or
equivalently, to the density fluctuations. These fluctuations are
themselves related to the compressibility and diverge at the critical
point~\cite{Pathria96}. As the density fluctuations diverge near the
phase transition, the neutrino opacity may increase
significantly. This could have a dramatic effect on present models of
stellar collapse. So far one dimensional simulations with the most 
sophisticated treatment of neutrino transport have not 
exploded~\cite{Bur03_PRL90}.

In the opposite $q\rightarrow \infty$ limit, the neutrino wavelength
is much shorter than the interparticle separation and the neutrino
resolves one nucleon at a time. This limit corresponds to quasielastic 
scattering where the cross section per nucleon in the medium is the 
same as in free space. Thus, the coherence disappears and
\begin{equation}
 S(q\rightarrow\infty)=1 \;.
 \label{slarge}
\end{equation}

In Monte Carlo as well as in molecular dynamics simulations it is
convenient to compute the static structure factor from 
the neutron-neutron correlation function $g(r)$. Indeed, the static
structure factor is obtained from the Fourier transform of the
two-neutron correlation function. That is,
\begin{equation}
 S({\bf q})=1+\rho_n\int d^3r \Big(g({\bf r})-1\Big) 
            \exp(i{\bf q}\cdot {\bf r}) \;.
 \label{sq4}
\end{equation}
The convenience of the two-neutron correlation function stems from 
the fact that it measures spatial correlations that may be easily
measured during the simulations. It is defined as follows:
\begin{equation}
 g({\bf r})=\frac{1}{N\rho_n} \sum_{i\neq j}^{N}
      \langle\Psi_{0}|\delta({\bf r}- {\bf r}_{ij})
      |\Psi_{0}\rangle \;,
 \label{gr}
\end{equation}
where $\rho_{n}\!\equiv\!N/V$ is the average neutron density. 
Operationally, the correlation function is measured by pausing
the simulation to compute the number of neutron pairs separated 
by a distance $|{\bf r}|$. Note that the two-neutron correlation 
function is normalized to one at large distances
$g(r\rightarrow\infty)\!=\!1$; this corresponds to the average 
density of the medium.

\section{Results}
\label{results}

In this section results are presented for a variety of neutron-rich
matter observables over a wide range of densities. Our goal is to
understand the evolution of the system with density. From the
low-density phase of isolated nuclei, through the complex pasta phase,
to uniform matter at high densities. All the results in this section
have been obtained with an electron fraction and a temperature fixed
at $Y_e\!=\!0.2$ and $T\!=\!1$ MeV, respectively. In core-collapse
supernova the electron fraction starts near $Y_e\!=\!0.5$ and drops as
electron capture proceeds. In a neutron star $Y_e$ is small---of the
order of $0.1$---as determined by beta equilibrium and the nuclear
symmetry energy (a stiff symmetry energy favors larger values for
$Y_e$~\cite{Hor02_PRC66}). Thus, the value of $Y_e\!=\!0.2$ adopted
here is representative of neutron-rich matter.

There are limitations in our simple semiclassical model at both low
and high temperatures. At very low temperatures the system will
solidify while at high temperatures the model may not calculate
accurately the free-energy difference between the liquid and the
vapor. As a result the pasta may melt at a somewhat too low of a
temperature. Results are thus presented for only $T\!=\!1$ MeV where
the model gives realistic results. Recall that our model reproduces
both the saturation density and binding energy of nuclear matter, and
the long-range Coulomb repulsion between clusters. Therefore, a good
description of the clustering should be expected.

The simulations start at the low baryon density of
$\rho\!=\!0.01$~fm$^{-3}$ (which corresponds approximately to 
$2\!\times\!10^{13}$~g/cm$^3$). This is about $1/15$ of nuclear-matter 
saturation density. At this low density the most time consuming part 
of the simulation is preparing appropriate initial conditions. This is
because the Coulomb barrier greatly hinders the motion of protons into
and out of the clusters. The Coulomb barrier becomes an even greater 
challenge at lower densities. Thus, no effort has been made to 
simulate densities below $0.01$~fm$^{-3}$.

In Ref.~\cite{Hor04_PRC69} results were presented for Monte Carlo
simulations with $A\!=\!4,000$ nucleons. Unfortunately, finite-size
effects make it difficult to calculate $S(q)$ accurately at small
momentum transfers from these ``small'' simulations. Further, at a
baryon density $\rho\!=\!0.01$~fm$^{-3}$ each side of the simulation
volume has a length of approximately $L\!=\!75$~fm. This is inadequate
as the wavelength of a 10~MeV neutrino is close to $120$~fm.
Therefore, in the present simulations the number of particles has been
increased by a full of order of magnitude to $A=40,000$ nucleons. This
results in a simulation volume that has increased to a cube of
$L=158.7$~fm on a side. The molecular dynamics simulation starts with
the nucleons uniformly distributed throughout the simulation volume
and with a velocity profile corresponding to a $T\!=\!1$~MeV Boltzmann
distribution. The system is then evolved according to velocity-Verlet
algorithm using a time step of the order of $\Delta
t\!=\!2$~fm/c. Velocity-Verlet is an energy conserving algorithm and
with this time step the total energy of the system is conserved to one
part in $10^5$. However, in order to preserve the temperature fixed at
$T\!=\!1$~MeV, the velocities of all the particles must be
continuously rescaled so that the kinetic energy per particle stays
approximately fixed at $3k_{\rm B}T/2$. For some excellent references
to molecular dynamics simulations we refer the reader
to~\cite{Erc97_WWW,Ves01_Kluwer,All03_Oxford}.

In an attempt to speed up equilibration, the temperature of the system
was occasionally raised during the evolution to $1.5-2$~MeV.  This
could aid the system move away from local minimum, as is
conventionally done with simulated annealing. Equilibration is checked
by monitoring the time dependence of the two-neutron correlation
function $g(r)$ and the static structure factor $S(q)$. The peak in
$S(q)$ was observed to grow slowly with time as the cluster size
increased. Changes became exceedingly slow after evolving the system
for a long total time of $1,287,000$~fm/c. Although long, further
changes in $S(q)$ over much larger time scales can not be ruled
out. This suggests that our $\rho=0.01$~fm$^{-3}$ results may include
a systematic error due to the long---but finite---equilibration
time. Fortunately, equilibrium seems to be reached much faster at
higher densities so that the slow equilibration probably
ceases to be a problem at these densities.

Results for two observables---the potential energy per particle and 
the pressure---are displayed in Table~\ref{Table2} as a function of 
density. Also included in the table are the number of nucleons and 
the total evolution time for each density. Note that the pressure of 
the system is computed from the {\it virial equation} as 
follows~\cite{Erc97_WWW}:
\begin{equation}
 P=\rho \left[k_{\rm B}T - \frac{1}{3A}
  \Big\langle 
    \sum_{i<j}r_{ij} \frac{dV}{dr}\Big|_{r_{ij}}
  \Big\rangle \right] \;.
 \label{Pressure}
\end{equation}
In the case of non-interacting particles, Eq.~(\ref{Pressure}) 
reduces to the well-known equation of state of a classical ideal 
gas. Thus, the second term in the above equation reflects the
modifications to the ideal-gas law due to the interactions.

\begin{table}[ht]
\caption{molecular dynamics simulation results. Here $\rho$ 
         (in fm$^{-3}$) is the baryon density, $A$ is the
         baryon number, $t_{\rm f}$ (in fm/c) is the total 
	 evolution time, $V/A$ (in MeV) is the potential 
	 energy per particle, $P$ (in MeV/fm$^{3}$) is the
	 pressure, and $S(0)$ is the approximate value of
	 the static structure factor at $q\!=\!0$ computed
	 as in Eq.~(\ref{s0}).}
 \begin{ruledtabular}
 \begin{tabular}{cccccc}
  $\rho$ & $A$ & $t_{\rm f}$ &$V/A$ & $P$ & $S(0)$ \\ 
  \hline
   $0.010$ & $\phantom{0}40,000$ & $1,287,000$ & 
   $-5.377(1)$ & $6.9\!\times\!10^{-3}$ & $0.790$ \\
   $0.025$ & $100,000$ & $\phantom{0,0}52,000$ & $-5.145(1)$ & 
   $3.2\!\times\!10^{-2}$ & $0.344$ \\
   $0.050$ & $100,000$ & $\phantom{0,0}28,000$ & $-4.463(1)$ & 
   $0.13$ & $0.139$ \\
   $0.075$ & $\phantom{0}40,000$ & $\phantom{0,0}60,000$ & 
   $-3.686(1)$ & $0.33$ & \ $0.077$ 
 \label{Table2}
 \end{tabular}
\end{ruledtabular}
\end{table}

\begin{figure}[ht]
\begin{center}
\includegraphics[width=2.75in,angle=270,clip=false]{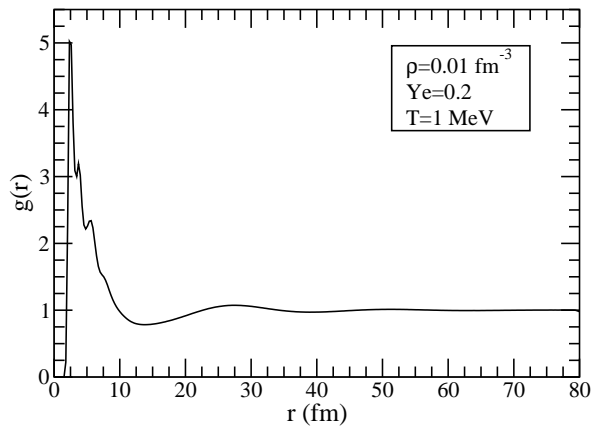}
\caption{Neutron-neutron correlation function $g(r)$ at a density
of $\rho\!=\!0.01$~fm$^{-3}$, an electron fraction of $Y_e\!=\!0.2$, 
and a temperature of $T\!=\!1$ MeV. This from a simulation with 
40,000 nucleons.}
\label{Fig1}
\end{center}
\end{figure}
   
The neutron-neutron correlation function $g(r)$ at a baryon density of
$\rho\!=\!0.01$~fm$^{-3}$ is shown in Fig.~\ref{Fig1}. The two-neutron
correlation function measures the probability of finding a pair of 
neutrons separated by a fixed distance $r$. A large broad peak is 
observed in $g(r)$ in the 2-10~fm region; the lack of 
neutrons with a relative distance of less than $2$~fm is due to the
hard core of the potential. The sharper sub-peaks contained in this
structure reflect neutron-neutron correlations (nearest neighbors,
next-to-nearest neighbors, and so on) within a single cluster. The
Coulomb repulsion among protons prevents the clusters from growing
arbitrarily large and keeps them apart. The dip in $g(r)$ at
$r\!\simeq\!10$~fm is a result of the Coulomb repulsion between
clusters. Finally, the small broad peaks near 25-30, 50, and 75~fm
reflect correlations among the different clusters.

\begin{figure}[ht]
\begin{center}
\includegraphics[width=2.75in,angle=270,clip=false]{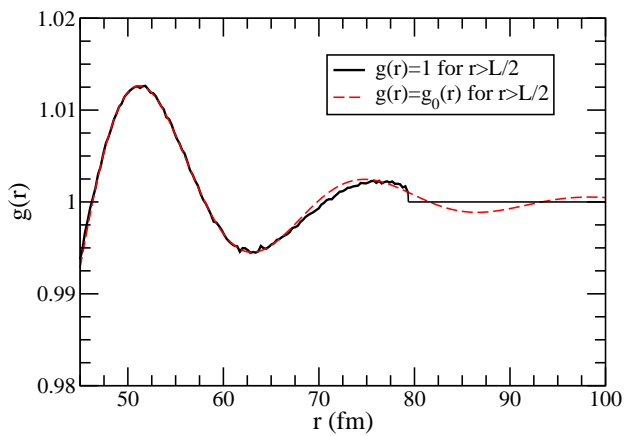}
\caption{(Color online) The large-$r$ behavior of the two-neutron
         correlation function displayed in Fig.~\ref{Fig1}. Note
	 the expanded y-scale. Also shown (red dashed line) is the 
	 analytic fit to $g(r)$ according to Eq.(\ref{gfit}).}
\label{Fig2}
\end{center}
\end{figure}

Figure~\ref{Fig2} displays an enlargement of the neutron-neutron
correlation function for large values of $r$.  Finite-size effects
lead to an abrupt drop in $g(r)$ at $r\!=\!L/2\!\sim\!80$~fm (not
shown). To ensure a reliable estimate of its Fourier transform---and
correspondingly of the static structure factor $S(q)$---one must
extrapolate $g(r)$ to the region $r\!>\!L/2$. To do so, an analytic
function of the following form is fitted to $g(r)$:
\begin{equation}
  g_0(r)=A_0 e^{-\alpha_0 r} \cos (k_0 r + \delta_0) + 1 \;.
  \label{gfit}
\end{equation}
The constants $A_0$, $\alpha_0$, $k_0$, and $\delta_0$ are obtained 
from a fit to the large-$r$ behavior of the neutron-neutron
correlation function. The result of this fit is indicated by the 
red dashed line in Fig.~\ref{Fig2}.

\begin{figure}[ht]
\begin{center}
\includegraphics[width=2.75in,angle=0,clip=false]{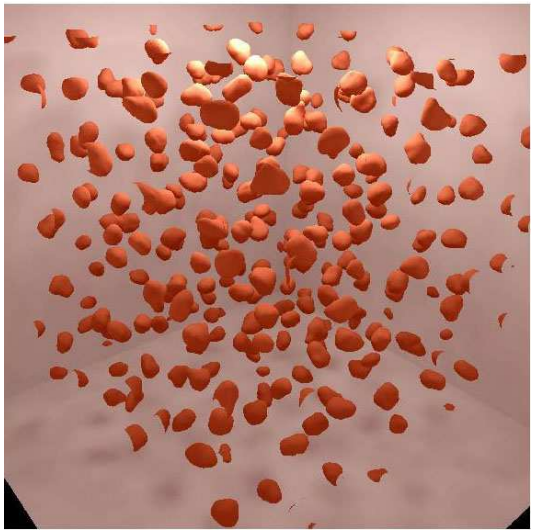}
\caption{(Color online) The 0.03 fm$^{-3}$ proton density isosurface for one configuration of 40,000 nucleons at a density of 0.01 fm$^{-3}$.  The simulation volume is a cube 159 fm on a side.}
\label{Fig3a}
\end{center}
\end{figure}

Figure~\ref{Fig3a} shows the 0.03 fm$^{-3}$ isosurface of the proton density for one configuration of 40,000 nucleons at a density of $0.01$ fm$^{-3}$.  All of the protons and most of the neutrons are clustered into nuclei.  We discuss the size of these nuclei in Section \ref{cluster_model}.  There is also a low density neutron gas between the clusters which is not shown.

\begin{figure}[ht]
\begin{center}
\includegraphics[width=2.75in,angle=270,clip=false]{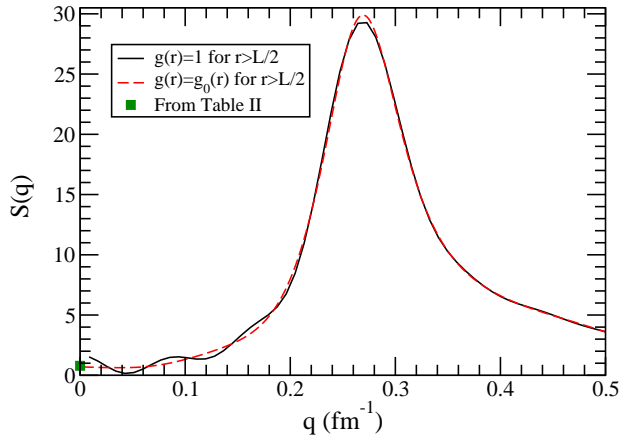}
\caption{(Color online) Static structure factor $S(q)$ at a density of
$\rho\!=\!0.01$ fm$^{-3}$, an electron fraction of $Y_e\!=\!0.2$, and
a temperature of $T\!=\!1$ MeV. The black solid line assumes
$g(r)\!=\!1$ for $r\!>\!L/2$ while the red dashed line includes an
analytic extrapolation for $g(r)$ for $r\!\agt\!L/2$. The green solid
square is the value of $S(0)$ from Eq.~(\ref{s0}) and
Table~\ref{Table2}.}
\label{Fig3}
\end{center}
\end{figure}

The static structure factor $S(q)$ may now be calculated from the 
Fourier transform of $g(r)$ [see Eq.~(\ref{sq4})]. That is,
\begin{equation}
 S(q)=1+4\pi\rho_{n}\int_{0}^{\infty}\frac{\sin(qr)}{qr}
        (g(r)-1)r^{2}dr \;. 
 \label{sq5}
\end{equation}
In Fig.~\ref{Fig3} the static structure factor obtained by 
using the above extrapolation ({\it i.e.,} with $g_0(r)$ 
for $r\!\agt\!L/2$) is displayed with the red dashed line. 
In an earlier publication the static structure factor was 
calculated directly from the simulation results assuming 
$g(r)\!=\!1$ for $r\!>\!L/2$~\cite{Hor04_PRC69}. This 
result is also shown for comparison (black solid line).
The improvement in the low-momentum transfer behavior of 
$S(q)$ is clearly evident. This is important as the value 
of the static structure factor at zero-momentum transfer 
$S(q\!=\!0)$ monitors density fluctuations in the system 
and these may be indicative of a phase transition.

In the limit of zero-momentum transfer the static structure factor 
$S(q\!=\!0)$ is directly related to the isothermal compressibility.
That is~\cite{Pathria96}, 
\begin{equation}
 S(q\!=\!0)=\rho\,k_{\rm B}T\,{\cal K}_{T}
           =k_{\rm B}T\left(\frac{\partial{P}}
              {\partial{\rho}}\right)^{-1}_{T} \;,
 \label{sq0}
\end{equation}
where the isothermal compressibility is given by
\begin{equation}
 {\cal K}_{T}^{-1}=-V \left(\frac{\partial{P}}
                     {\partial{V}}\right)_{T}
                  =\rho\left(\frac{\partial{P}}
                   {\partial{\rho}}\right)_{T} \;.
 \label{compress}
\end{equation} 
For a classical ideal gas ({\it i.e.,} $P\!=\!\rho k_{\rm B}T$) the 
isothermal compressibility reduces to 
${\cal K}_{T}^{-1}\!=\!\rho k_{\rm B}T$ and $S(q\!=\!0)\!=\!1$.
As expected, in the absence of interactions there are no spatial 
correlations among the particles. Assuming now that as
$q\!\rightarrow\!0$ the fluctuations in the neutron density are 
proportional to the corresponding fluctuations in the baryon
density, one obtains for the static structure factor per neutron
\begin{equation}
  S(q\!=\!0) = \frac{N}{A}k_{\rm B}T
  \left(\frac{\partial{P}}{\partial{\rho}}\right)^{-1} \;.
 \label{sq0neutrons}
\end{equation}
The derivative of the pressure with respect to the baryon density has
not been directly calculated in the simulations. However, the pressure
has been computed at various densities and has been tabulated in
Table~\ref{Table2}. These values can be approximated by a simple
fit of the form:
\begin{equation}
 P(\rho)=(0.611)\rho+(228.131)\rho^{\alpha+1} 
 \quad (\alpha=1.583)\;,
 \label{Pfit}
\end{equation}
with the pressure expressed in units of MeV/fm$^3$ and the density
in fm$^{-3}$. This yields,
\begin{equation}
 S(q\!=\!0)\approx \frac{1.358\!\times\!10^{-3}}
           {\left(1.037\!\times\!10^{-3}+\rho^{\alpha}\right)}\;.
 \label{s0}
\end{equation}
These approximate values for $S(q\!=\!0)$ have been reported
in Table~\ref{Table2}. For comparison, they have also been added 
(with a green solid square) to the various figures for which $S(q)$ 
was directly computed from the Fourier transform of the two-neutron
correlation function (see Figs.~\ref{Fig3},~\ref{Fig6},
and~\ref{Fig9}). Note that there is good agreement between the two
approaches.
 
At low momentum transfers the static structure factor is small because
of ion screening. Coulomb correlations---which both hinder the growth
of clusters and keep them well separated---screen the weak charge of
the clusters thereby reducing $S(q)$. Further, a large peak is seen in
$S(q)$ for $q\!\simeq\!0.25$~fm$^{-1}$. This corresponds to the
coherent scattering from all of the neutrons in a cluster. As we will
show in the next section, this peak reproduces coherent
neutrino-nucleus elastic scattering at low densities. Finally, the
static structure factor decreases for $q\!>\!0.3$~fm$^{-1}$. This is
due to the form factor of a cluster. As the momentum-transfer
increases, the neutrino can no longer scatter coherently form all the
neutrons because of the size of the cluster is larger than the
neutrino wavelength. All these features will be discussed in greater
detail in the next section.

\begin{figure}[ht]
\begin{center}
\includegraphics[width=2.75in, angle=270,clip=false]{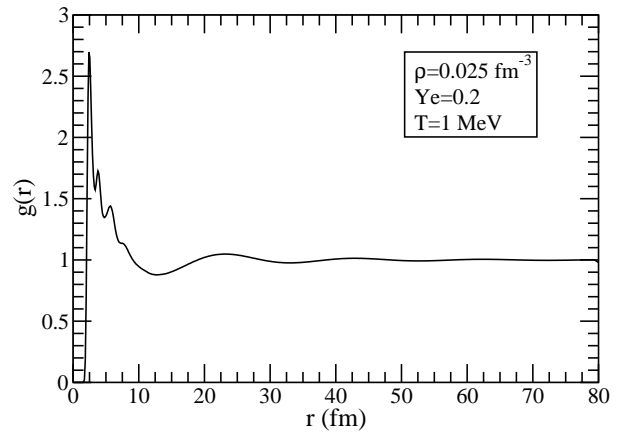}
\caption{Neutron-neutron correlation function $g(r)$ at a density
of $\rho\!=\!0.025$~fm$^{-3}$, an electron fraction of $Y_e\!=\!0.2$, 
and a temperature of $T\!=\!1$ MeV. This from a simulation with 
100,000 nucleons.}
\label{Fig4}
\end{center}
\end{figure}

\begin{figure}[ht]
\begin{center}
\includegraphics[width=2.75in,angle=270,clip=false]{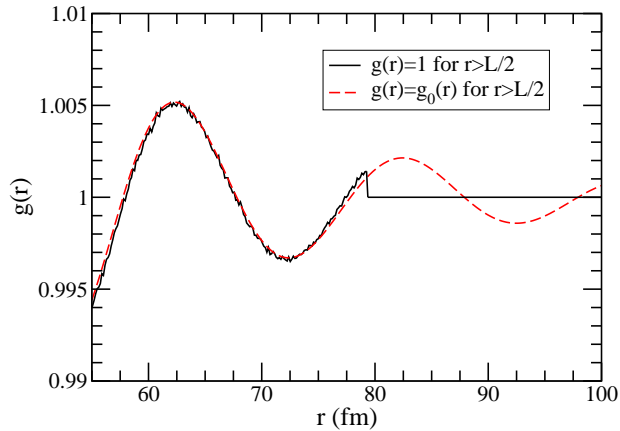}
\caption{(Color online) The large-$r$ behavior of the two-neutron
         correlation function displayed in Fig.~\ref{Fig4}. Also
	 shown (red dashed line) is the analytic fit to $g(r)$.}
\label{Fig5}
\end{center}
\end{figure}

Next, simulation results are presented at the higher density of
$\rho\!=\!0.025$~fm$^{-3}$. At this density it becomes much easier to
equilibrate the system as protons have shorter distances to move over
the Coulomb barriers. In order to minimize finite-size effects more
nucleons---a total of $A\!=\!100,000$---are used for these
simulations. The two-neutron correlation function is shown in
Fig.~\ref{Fig4}, with its behavior at large distances amplified in
Fig.~\ref{Fig5}. Note that the calculation for $g(r)$, which
proceeds by histograming distances between the $N(N\!-\!1)/2$ pairs 
of neutrons, is now considerably more time consuming. The resulting
static structure factor is shown in Fig.~\ref{Fig6}. The peak in 
$S(q)$ has now moved to larger $q$ because of the shorter distance 
between clusters.

\begin{figure}[ht]
\begin{center}
\includegraphics[width=2.75in,angle=0,clip=false]{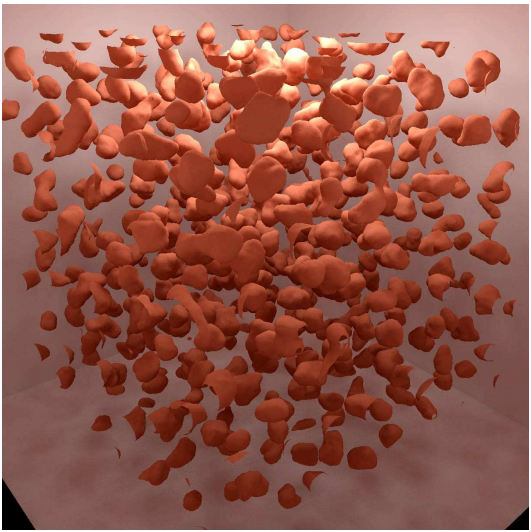}
\caption{(Color online) The 0.03 fm$^{-3}$ proton density isosurface for one configuration of 100,000 nucleons at a density of 0.025 fm$^{-3}$.  The simulation volume is a cube 159 fm on a side.}
\label{Fig6a}
\end{center}
\end{figure}

Figure~\ref{Fig6a} shows the 0.03 fm$^{-3}$ isosurface of the proton density for one configuration of 100,000 nucleons at a density of $0.025$ fm$^{-3}$.  All of the protons and most of the neutrons are clustered into nuclei.  The size of these nuclei is now larger than at a density of $0.01$ fm$^{-3}$ as discussed in Section \ref{cluster_model}.  There is also a low density neutron gas between the clusters which is not shown. 

\begin{figure}[ht]
\begin{center}
\includegraphics[width=2.75in,angle=270,clip=false]{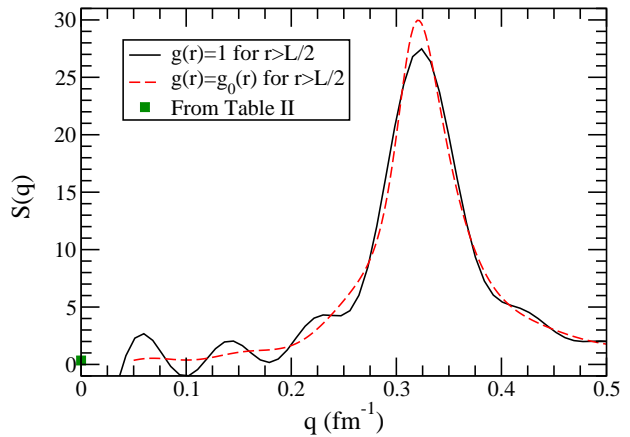}
\caption{(Color online) Static structure factor $S(q)$ at a density of
$\rho\!=\!0.025$~fm$^{-3}$, an electron fraction of $Y_e\!=\!0.2$, and
a temperature of $T\!=\!1$ MeV. The black solid line assumes
$g(r)\!=\!1$ for $r\!>\!L/2$ while the red dashed line includes an
analytic extrapolation for $g(r)$ for $r\!\agt\!L/2$. The green solid
square is the value of $S(0)$ from Eq.~(\ref{s0}) and
Table~\ref{Table2}.}
\label{Fig6}
\end{center}
\end{figure}

\begin{figure}[ht]
\begin{center}
\includegraphics[width=2.75in,angle=270,clip=false]{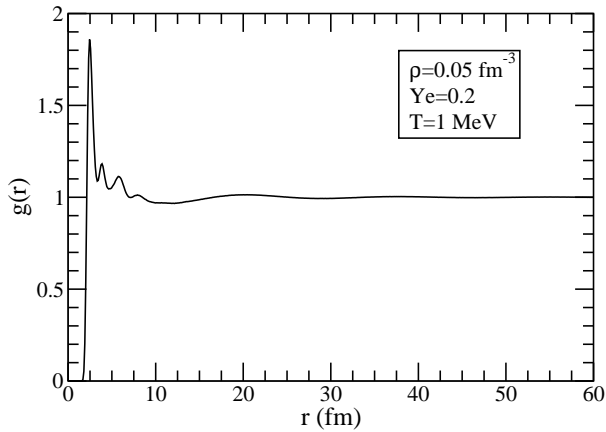}
\caption{Neutron-neutron correlation function $g(r)$ at a density
of $\rho\!=\!0.05$~fm$^{-3}$, an electron fraction of $Y_e\!=\!0.2$, 
and a temperature of $T\!=\!1$ MeV. This is from a simulation with 
100,000 nucleons.}
\label{Fig7}
\end{center}
\end{figure}

\begin{figure}[ht]
\begin{center}
\includegraphics[width=2.75in,angle=270,clip=false]{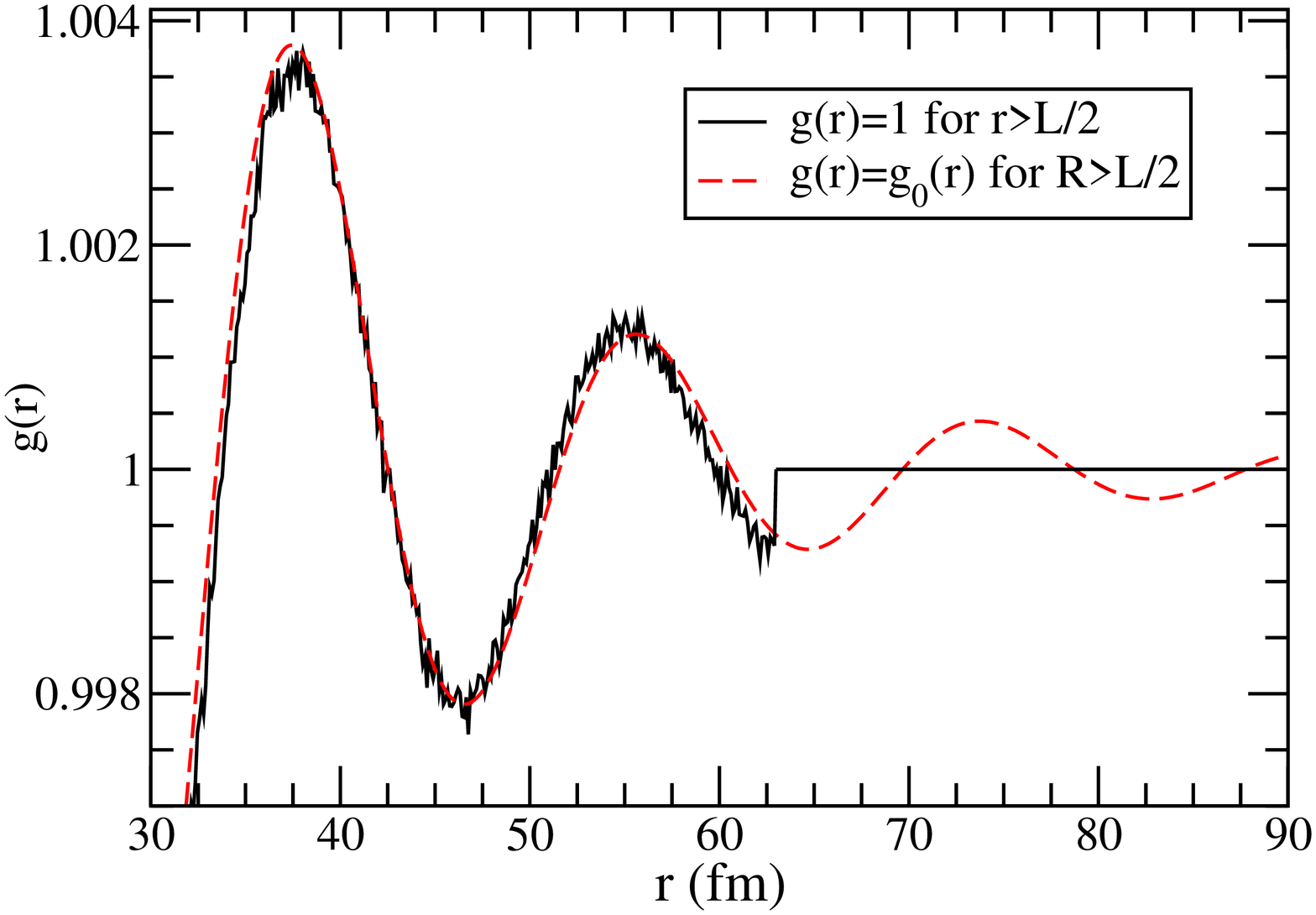}
\caption{(Color online) The large-$r$ behavior of the two-neutron
         correlation function displayed in Fig.~\ref{Fig7}. Also
	 shown (red dashed line) is the analytic fit to $g(r)$.}
\label{Fig8}
\end{center}
\end{figure}

Simulations have also been performed at a density of
$\rho\!=\!0.05$~fm$^{-3}$ using $A\!=\!100,000$ nucleons.  The
two-neutron correlation function, together with its amplification at
large values of $r$, are shown Figs.~\ref{Fig7} and~\ref{Fig8},
respectively. The corresponding static structure factor is displayed
Fig.~\ref{Fig9}. Note the significant improvement in the behavior of
$S(q)$ at low momentum transfers as the sharp cutoff in $g(r)$ is
removed in favor of a smooth extrapolation [see Eq.~(\ref{gfit})].  As
the density increases, and thus the separation between clusters
decreases, the peak in $S(q)$ continues to move to higher $q$. 
However, the peak value of $S(q)$ has now been reduced because of
the increase in ion screening with density.

\begin{figure}[ht]
\begin{center}
\includegraphics[width=2.75in,angle=0,clip=false]{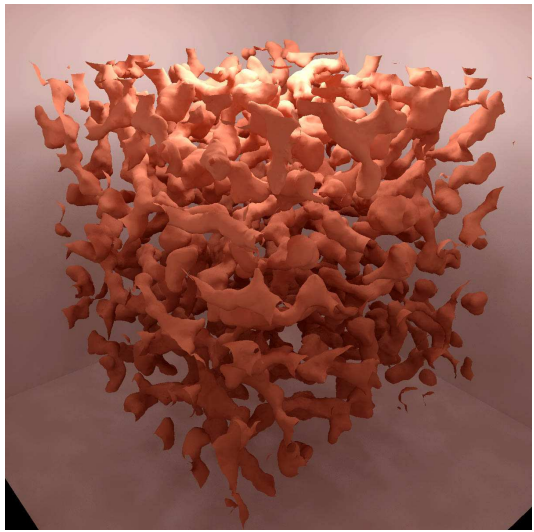}
\caption{(Color online) The 0.03 fm$^{-3}$ proton density isosurface for one configuration of 100,000 nucleons at a density of 0.05 fm$^{-3}$.  The simulation volume is a cube 126 fm on a side.}
\label{Fig9a}
\end{center}
\end{figure}

Figure~\ref{Fig9a} shows the 0.03 fm$^{-3}$ isosurface of the proton density for one configuration of 100,000 nucleons at a density of $0.05$ fm$^{-3}$.  The clusters are now seen to have very elongated shapes.  The low density neutron gas between these clusters is not shown. 

\begin{figure}[ht]
\begin{center}
\includegraphics[width=2.75in,angle=270,clip=false]{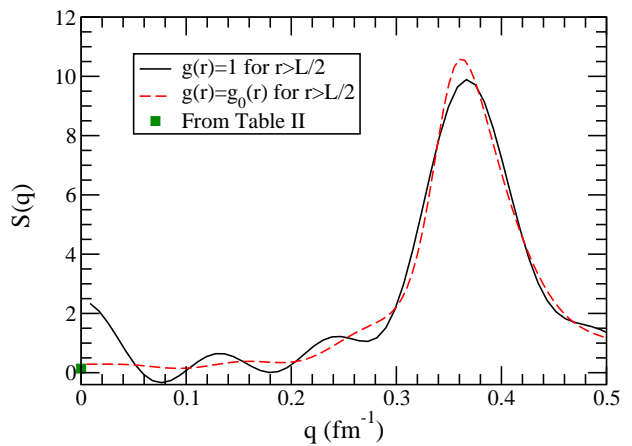}
\caption{(Color online) Static structure factor $S(q)$ at a density
of $\rho\!=\!0.05$~fm$^{-3}$, an electron fraction of $Y_e\!=\!0.2$, 
and a temperature of $T\!=\!1$ MeV. The black solid line assumes 
$g(r)\!=\!1$  for $r\!>\!L/2$ while the red dashed line includes an 
analytic extrapolation for $g(r)$ for $r\!\agt\!L/2$. The green solid 
square is the value of $S(0)$ from Eq.~(\ref{s0}) and 
Table~\ref{Table2}.}
\label{Fig9}
\end{center}
\end{figure}

\begin{figure}[ht]
\begin{center}
\includegraphics[width=2.75in,angle=270,clip=false]{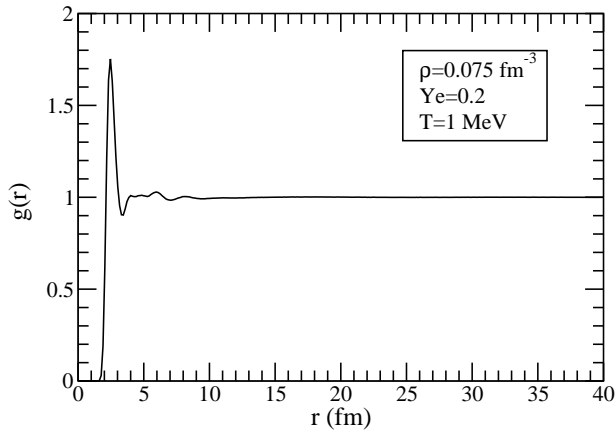}
\caption{Neutron-neutron correlation function $g(r)$ at a density
of $\rho\!=\!0.075$~fm$^{-3}$, an electron fraction of $Y_e\!=\!0.2$, 
and a temperature of $T\!=\!1$ MeV. This from a simulation with 
40,000 nucleons.}
\label{Fig10}
\end{center}
\end{figure}

\begin{figure}[ht]
\begin{center}
\includegraphics[width=2.75in,angle=270,clip=false]{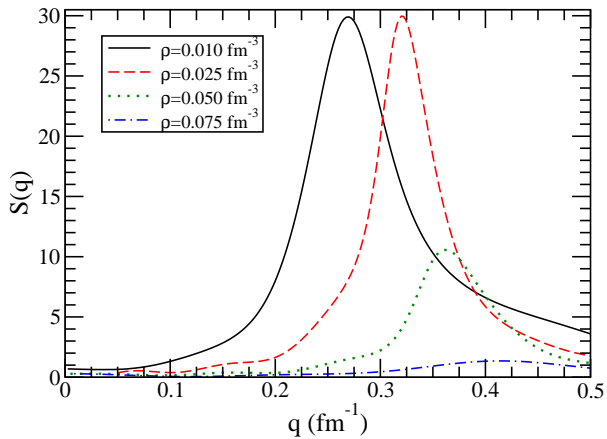}
\caption{(Color online) Static structure factor $S(q)$ for a 
variety of densities at an electron fraction of $Y_e\!=\!0.2$ 
and a temperature of $T\!=\!1$ MeV.  The black solid line is 
for a density of $\rho\!=\!0.01$~fm$^{-3}$, while the red 
dashed line is for $\rho\!=\!0.025$~fm$^{-3}$, the green dotted  
line for $\rho\!=\!0.05$~fm$^{-3}$, and the blue dot-dashed
line for $\rho\!=\!0.075$~fm$^{-3}$.}
\label{Fig11}
\end{center}
\end{figure}

\begin{figure}[ht]
\begin{center}
\includegraphics[width=2.75in,angle=270,clip=false]{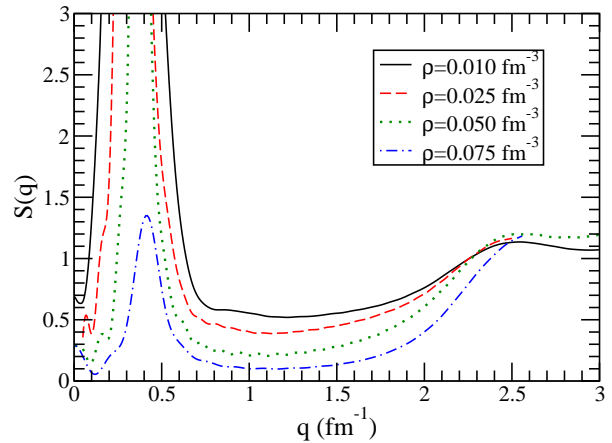}
\caption{(Color online) The large-$q$ behavior of the static
         structure function displayed in Fig.~\ref{Fig11}.}
\label{Fig12}
\end{center}
\end{figure}

We conclude this section by presenting results for the two-neutron
correlation function at a density of $\rho\!=\!0.075$~fm$^{-3}$ 
using a total of $A=40,000$ nucleons in Fig.~\ref{Fig10}. Note that
this is the largest density considered in this work. At this density
the clusters have been ``melted'' and the system has evolved into a
uniform phase. In Fig.~\ref{Fig11} results for the static structure
factor at this density are compared with the corresponding results at
the lower densities.  There is no longer evidence for a large peak in 
$S(q)$ in the uniform system as a consequence of the complete loss in
coherence. Finally, Fig.~\ref{Fig12} shows $S(q)$ at large momentum
transfers. One observes that the static structure factor decreases
with increasing density in the intermediate $q$-region before
approaching the value of one (as it must) at high $q$.

In the next section we will compare these ``complete'' results with
conventional approaches that model the system as a collection of
strongly-correlated, neutron-rich nuclei plus a neutron gas. In this
approach the peak observed in the static structure factor is
attributed to {\it neutrino-nucleus elastic scattering}. Note,
however, that the complete simulation results obtained in this 
section should remain valid even when these nuclear models break 
down.

\section{Cluster Model}
\label{cluster_model}

In this section a cluster model is developed with the goal of comparing
our simulation results from the previous section with commonly used
approaches. In the complete model employed earlier, trajectories for all
the nucleons were calculated from molecular dynamics simulations
and these were used to compute directly the two-neutron correlation
function and its corresponding static structure factor. One of the main
virtues of such an approach is that there is no need to decide if a given
nucleon is part of a cluster or part of the background vapor.
Nevertheless, in this section a clustering algorithm is constructed with
the aim of assigning nucleons to clusters.  In this way one can compare
the inferred composition extracted from our simulations with many nuclear
statistical equilibrium (NSE) models that describe the system as a
collection of nuclei and free nucleons. In this way one can then compare
the static structure factor extracted from the complete simulations
with that calculated in these NSE models.

\subsection{Clustering Algorithm}

The clustering algorithm implemented in this section assigns a nucleon
to a cluster if it is within a distance $R_{C}$ of at least one other
nucleon in the cluster.  In practice, one stars with a given nucleon
and searches for all of its ``neighbors'', namely, all other nucleons
contained within a sphere of radius $R_C$.  Next, one repeats the same
procedure for all of its neighbors until no new neighbors are
found. This procedure divides a fixed configuration of nucleons into a
collection of various mass clusters ({\i.e.,} ``nuclei'').

\begin{figure}[ht]
\begin{center}
\includegraphics[width=2.75in,angle=270,clip=false]{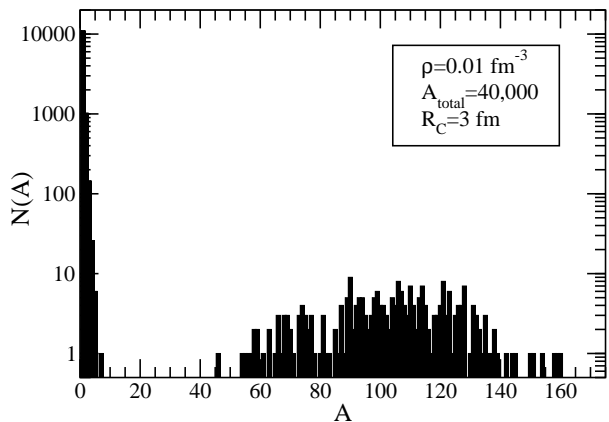}
\caption{Number of clusters of atomic mass $A$ for one configuration
of 40,000 nucleons at a density of $\rho\!=\!0.01$ fm$^{-3}$.  Note
that this is a linear-log plot.}
\label{Fig13}
\end{center}
\end{figure}

To illustrate this procedure the final nucleon configuration of the
complete simulation of the previous section at a density of
$\rho=0.01$~fm$^{-3}$ is selected after the system has evolved for 
a total time of $t_{\rm f}\!=\!1,287,000$~fm/c. Having selected a 
cutoff radius of $R_C\!=\!3$~fm, one finds that the 40,000 nucleons 
in the system are divided in the following way: a) 11,062 free
neutrons, b) no free protons, c) a few light nuclei with $A\!<\!8$, 
and d) a collections of heavy nuclei with a mass distribution of
$50\!\lesssim\!A\!\lesssim\!160$. The mass-weighted average of all
clusters with $A\!>\!2$ is equal to $\langle A \rangle\!=\!99$. This
distribution of clusters is displayed in Fig.~\ref{Fig13} and listed
in Table~\ref{Table3}.

\begin{table}
\caption{Distribution of cluster as a function of the cutoff radius
$R_C$ at a density $\rho\!=\!0.01$ fm$^{-3}$ for a system of 40,000
nucleons.  The number of free neutrons is denoted by $N(A\!=\!1)$, the
average mass of all $A\!>\!2$ clusters by $\langle A\rangle$, and the
size of the largest cluster by $A_{\rm max}$.}
\begin{ruledtabular}
\begin{tabular}{cccc}
 $R_C$ (fm) & $N(A\!=\!1)$ & $\langle A\rangle$ & $A_{\rm max}$ \\
  \hline
  2.0 & 20,585 &  46.20 &    118 \\
  2.5 & 14,233 & 102.29 &    155 \\
  3.0 & 11,062 &  98.75 &    160 \\
  3.5 &  7,856 &  96.17 &    254 \\
  4.0 &  5,131 & 100.97 &    261 \\
  4.5 &  2,888 & 149.84 &    513 \\
  5.0 &  1,427 & 14,069 & 23,189
\label{Table3}
\end{tabular}
\end{ruledtabular}
\end{table}

In Table~\ref{Table3} results are also displayed for values of the
cutoff radius in the 2-5~fm range. Note that the average
cluster mass $\langle A\rangle$ appears remarkably constant for
$2.5\!\leq\!R_C\!\leq\!4$~fm.  This suggest that any value of $R_C$
within this range should give similar results. However, if $R_C$ is
chosen too large, for example $R_C\!=\!5$~fm, then most of the
nucleons become part of one single giant cluster.

\begin{figure}[ht]
\begin{center}
\includegraphics[width=2.75in,angle=270,clip=false]{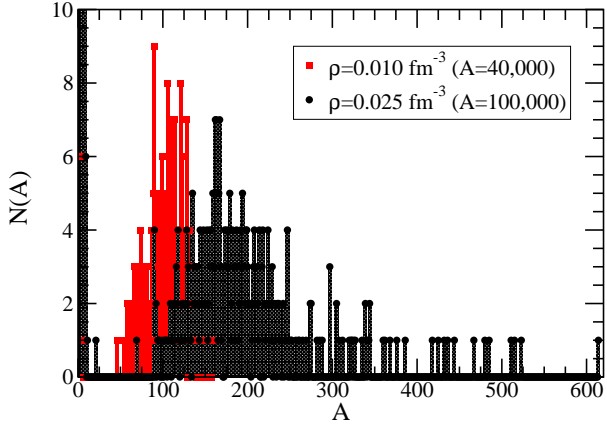}
\caption{(Color online) Number of clusters of atomic mass $A$ for 
one configuration of 100,000 nucleons at a density of 
$\rho\!=\!0.025$~fm$^{-3}$ (black hatched line). Also shown for 
comparison is the number of clusters (from Fig.~\ref{Fig13}) at 
$\rho\!=\!0.01$~fm$^{-3}$ (red solid line). Note that both 
scales are now linear.}
\label{Fig14}
\end{center}
\end{figure}

Similar results for a density of $\rho\!=\!0.025$ fm$^{-3}$ are presented in 
Table~\ref{Table4}. This distribution is extracted from the final
configuration of 100,000 nucleons obtained after a total evolution
time of $t_{\rm f}\!=\!52,000$~fm/c. Using a cutoff radius of $R_C=3$
fm, the 100,000 nucleons in the system are now divided into 14,549
free neutrons, no free protons, a few light nuclei, and a broad
collection of heavy nuclei with $A$ from about 80 to 614 nucleons.
Such a distribution is shown in Fig.~\ref{Fig14}. The average mass 
has now grown to $\langle A\rangle\!=\!199$. The mass of the heavy 
nuclei is seen to increase with density as shown in Fig. \ref{Fig14}.  
Finally, results at $\rho\!=\!0.05$ fm$^{-3}$ are presented in 
Table~\ref{Table5}. Now the density is so high that it is difficult 
to design a sensibly scheme to divide the system into clusters, see Fig. \ref{Fig9a}.  For example, even with a cutoff radius as small as $R_C\!=\!2.5$ fm, already 78,178 of the 100,000 nucleons become part of a single giant 
cluster.

\begin{table}
\caption{Distribution of cluster as a function of the cutoff radius
$R_C$ at a density $\rho\!=\!0.025$ fm$^{-3}$ for a system of 100,000
nucleons.  The number of free neutrons is denoted by $N(A\!=\!1)$, the
average mass of all $A\!>\!2$ clusters by $\langle A\rangle$, and the
size of the largest cluster by $A_{\rm max}$.}
\begin{ruledtabular}
\begin{tabular}{cccc}
 $R_C$ (fm) & $N(A\!=\!1)$ & $\langle A\rangle$ & $A_{\rm max}$ \\
 \hline
2.0 & 49,761 &  58.95 &      162 \\
2.5 & 28,234 & 167.40 &      423 \\
3.0 & 14,549 & 198.91 &      614 \\
3.5 &  5,187 & 59,411 &   75,003 
\label{Table4}
\end{tabular}
\end{ruledtabular}
\end{table}

\begin{table}
\caption{Distribution of cluster as a function of the cutoff radius
$R_C$ at a density $\rho\!=\!0.05$ fm$^{-3}$ for a system of 100,000
nucleons. The number of free neutrons is denoted by $N(A\!=\!1)$, the
average mass of all $A\!>\!2$ clusters by $\langle A\rangle$, and the
size of the largest cluster by $A_{\rm max}$.}
\begin{ruledtabular}
\begin{tabular}{cccc}
 $R_C$ (fm) & $N(A\!=\!1)$ & $\langle A\rangle$ & $A_{\rm max}$ \\
 \hline
2.0 & 47,239 &  56.86 &      308 \\
2.5 & 16,219 & 72,952 &   78,178 
\label{Table5}
\end{tabular}
\end{ruledtabular}
\end{table}

\subsection{Cluster Form Factors}

\begin{figure}[ht]
\begin{center}
\includegraphics[width=2.75in,angle=0,clip=false]{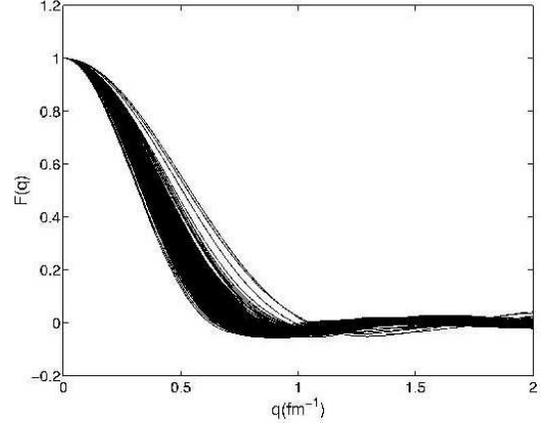}
\caption{Cluster form factor $F(q)$ as a function of the momentum 
transfer $q$ for all clusters with $A\!>\!10$ using a single 
configuration of 40,000 nucleons at a density of 
$\rho\!=\!0.01$ fm$^{-3}$.}
\label{Fig15}
\end{center}
\end{figure}

\begin{figure}[ht]
\begin{center}
\includegraphics[width=2.75in,angle=0,clip=false]{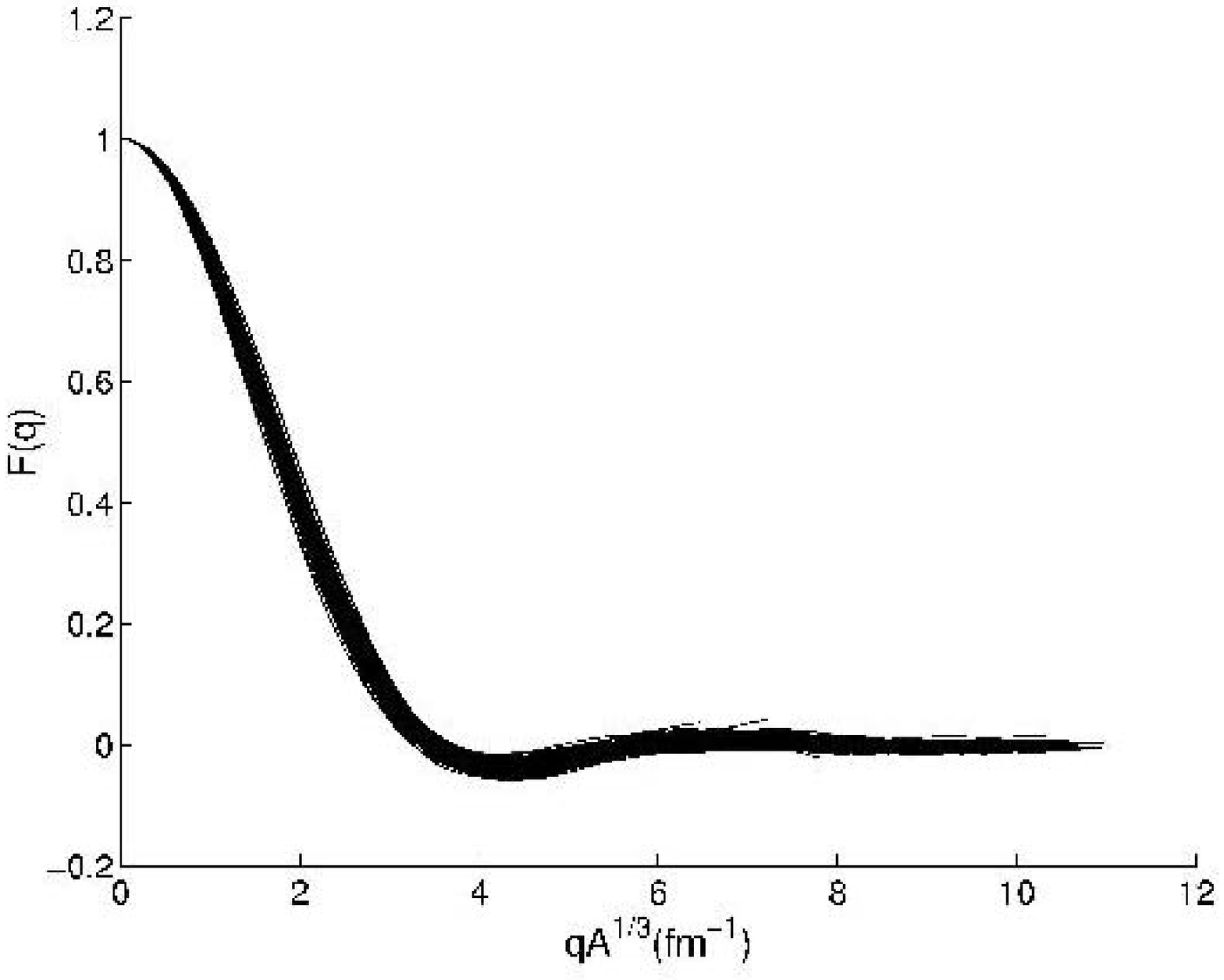}
\caption{Cluster form factor $F(q)$ as a function of the scaled 
momentum transfer $q A^{1/3}$ for all clusters with $A\!>\!10$ 
using a single configuration of 40,000 nucleons at a density of 
$\rho\!=\!0.01$ fm$^{-3}$.}
\label{Fig16}
\end{center}
\end{figure}

To describe coherent neutrino scattering from a single 
cluster, that is, {\it neutrino-nucleus elastic scattering},
one must calculate the elastic form factor for the cluster.
This is given by
\begin{equation}
  F(q)=\frac {1}{N} \sum_{n=1}^N 
       \frac{\sin(q r_n)}{q r_n} \;.
\label{F}
\end{equation}
Here the sum runs over the $N$ neutrons in the cluster and $r_n$ is
distance from the n$_{\rm th}$ neutron to the center of mass of the
$N$-neutron system. The form factor represents the Fourier transform
of the point neutron density and here, for simplicity, has been
averaged over the direction of the momentum transfer.  Note that the
elastic form factor is normalized so that $F(q\!=\!0)\!=\!1$. In
Fig.~\ref{Fig15} the elastic form factors of all clusters with
$A\!>\!10$ are displayed at a density of $\rho\!=\!0.01$~fm$^{-3}$ (a
cutoff radius of $R_C\!=\!3$~fm was selected).  The large spread in
the form factors reflects the many different sizes of the individual
clusters (see Fig.~\ref{Fig13}). Indeed, the root-mean-square (RMS)
radius of a cluster appears to scale approximately as
$A^{1/3}$. Therefore, in Fig.~\ref{Fig16} all of these form factors
are plotted but against a {\it scaled momentum transfer}
$qA^{1/3}$. Now all the (scaled) form factors fall in a fairly narrow
band suggesting that, while these neutron-rich clusters have different
radii, they all share a similar shape.

\begin{figure}[ht]
\begin{center}
\includegraphics[width=2.75in,angle=270,clip=false]{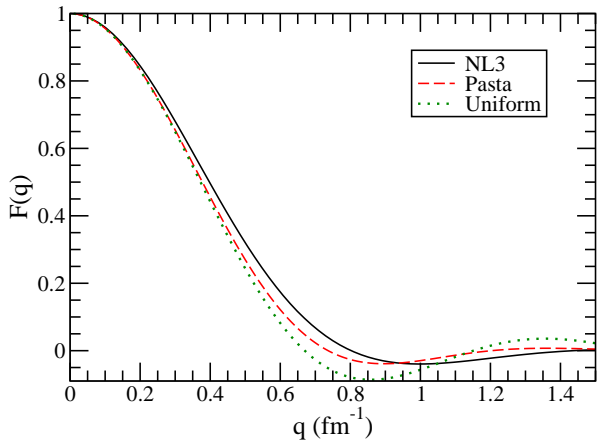}
\caption{(Color online) Cluster form factor $F(q)$ as a function 
of the momentum transfer $q$. The red dashed line represents the 
angle-averaged form factor for one cluster with $A\!=\!100$ and 
$Z\!=\!28$ from a simulation at a density of 
$\rho\!=\!0.01$~fm$^{-3}$. The dotted (green) line is the form 
factor of a uniform density sphere with the same root-mean-square 
radius [see Eqs.~(\ref{F0}) and~(\ref{sharpR})]. Finally, the solid 
line is the form factor of the very neutron-rich nucleus $^{98}$Ni 
calculated in a relativistic mean-field approximation with the 
NL3 interaction~\protect{\cite{Lal97_PRC55}}.}
\label{Fig17}
\end{center}
\end{figure}

In Fig.~\ref{Fig17} we display the form factor for a single
neutron-rich cluster with $A=100$ and $Z=28$ (``${}^{100}$Ni'') 
extracted from the simulation with a density of 
$\rho\!=\!0.01$ fm$^{-3}$.  Also shown in the figure (with a green 
dotted line) is the form factor $F_0(q)$ of a uniform neutron 
distribution with a sharp surface radius $R_n$ chosen to reproduce 
the RMS radius of the neutron-rich cluster $<r_n^2>^{1/2}$. It is
given by
\begin{equation}
 F_0(q)=3\frac{\sin(x)-x\cos(x)}{x^{3}} 
        \quad (x\equiv q R_n) \;,
\label{F0}
\end{equation}
with $R_n$ given by
\begin{equation}
  R_{n}=\sqrt{\frac{5}{3}}\langle r_{n}^{2}\rangle^{1/2}\;.
  \label{sharpR}
\end{equation}
Finally, Fig.~\ref{Fig17} also shows the neutron form factor of the
exotic, neutron-rich nucleus ${}^{98}$Ni calculated in a relativistic
mean-field approximation using the very successful NL3
interaction~\cite{Lal97_PRC55}. While the NL3 form factor has a
slightly smaller RMS radius, the overall agreement between all three
models is fairly good. Note that ${}^{98}$Ni (rather than
${}^{100}$Ni) was used in this calculation as it contains closed
protons and neutrons subshells. For this exotic nucleus the
$1h^{11/2}$ neutron orbit---responsible for magic number 82---is not
even bound.

The clusters generated in the simulations are neutron-rich nuclei 
with well developed neutron skins. Nuclei with neutron skins are
characterized by neutron radii that are larger than those for
the protons. Using the distribution of nuclei obtained with a
density of $\rho\!=\!0.01$~fm$^{-3}$, the following values are
obtained for average matter, proton, and neutron RMS radii 
respectively:
\begin{subequations}
 \begin{eqnarray}
  && \langle r^{2}\rangle^{1/2}=1.06\, A^{1/3}\, {\rm fm}\;, \\
  && \langle r_{p}^{2}\rangle^{1/2}=0.91\, A^{1/3}\, {\rm fm}\;, \\
  && \langle r_{n}^{2}\rangle^{1/2}=1.11\, A^{1/3}\, {\rm fm}\;.
 \end{eqnarray}
 \label{RMSradii}
\end{subequations}
Note that the sharp-surface radius of a uniform distribution with 
the same RMS radius is simply given by $(5/3)^{1/2}$ times these 
values [see Eq.~(\ref{sharpR})].

\subsection{Single Heavy Nucleus Approximation}

A number of approaches to dense matter, such as those using the
equation of state by Lattimer and Swesty~\cite{Lat92_NPA535}, model
the system as a collection of free neutrons plus a single
representative heavy nucleus. Occasionally, free protons and alpha
particles are also added to the system. To mimic this approach, a
model is constructed based on our earlier cluster results reported in
Tables~\ref{Table3} and~\ref{Table4} for a cutoff radius of
$R_C\!=\!3$~fm. For example, at a density of $\rho\!=\!0.01$~fm$^{-3}$
the system contains a mass fraction $X_n\!=\!0.28$ of free neutrons
and a mass fraction of $X_h\!=\!1-X_n\!=\!0.72$ for the single
representative heavy nucleus. According to the average mass reported
in Table~\ref{Table3}, a mass of $A\!=\!100$ is assigned to this
representative heavy nucleus. Conservation of charge constrains this
nucleus to have $Z\!\approx\!28$. Note that due to the presence of
free neutrons (but not free protons) the charge-to-mass ratio of the
heavy nucleus $Z/A\!=\!0.28$ slightly exceeds the electron fraction
$Y_e\!=\!0.2$ of the whole system.  The assumed composition of the
system at densities of $\rho\!=\!0.01$~fm$^{-3}$ and
$\rho\!=\!0.025$~fm$^{-3}$ is given in Table~\ref{Table6}.

\begin{table}
\caption{Composition of the system in the single-heavy-nucleus
approximation. The mass fraction of free neutrons is denoted by
is $X_n$ and that of heavy nuclei by $X_h$. The mass and charge
of the nuclei are given by $A$ and $Z$, respectively. Finally, 
the radii of the equivalent uniform proton and neutron 
distributions are denoted by $R_p$ and $R_n$, respectively.}
\begin{ruledtabular}
 \begin{tabular}{ccccccc}
  $\rho$ (fm$^{-3}$) & $X_n$ & $X_h$ & 
  A & Z & $R_p$ (fm) & $R_n$ (fm) \\
  \hline
  0.010 & 0.28 & 0.72 & 100 & 28 & 5.45 &   6.68  \\
  0.025 & 0.14 & 0.86 & 199 & 47 & 6.84 & \ 8.40 
 \label{Table6}
 \end{tabular}
\end{ruledtabular}
\end{table}

The heavy nuclei are assumed to interact exclusively via a screened
Coulomb interaction. Each nucleus is assumed to have a uniform charge
distribution $\rho_{\rm ch}$ that extends out to a radius $R_p$ chosen
to reproduce the proton RMS radius $\langle r_{p}^{2}\rangle^{1/2}$
given in Eq.~(\ref{RMSradii}). The Coulomb interaction between two 
such nuclei whose centers are separated by a distance $R$ is given by
 \begin{equation}
  V_C(R)=e^{2} \int d^3r\,\rho_{\rm ch}(r) 
               \int d^3r^{\prime}\,\rho_{ch}(r^\prime) 
               \frac {e^{-R_{\rm tot}/\lambda}}{R_{\rm tot}},
\label{Vc0}
\end{equation}
where $R_{\rm tot}\equiv|{\bf R}+{\bf r}-{\bf r}^{\prime}|$ and
$\lambda$ is the screening length fixed (as in Sec.~\ref{formalism})
at a constant value of $\lambda\!=\!10$~fm. In the limit that the
distance between nuclei is much larger than the nuclear RMS radius
({\it i.e.,} $R_{p}\!\ll\!R$) the above integral reduces to  
 \begin{eqnarray}
  && V_C(R) \simeq e^{2}\rho_{\rm ch}^{2}\frac{e^{-R/\lambda}}{R}
            \Big(\int d^3r\,e^{-r\cos\theta/\lambda}\Big)^{2} 
            \nonumber \\
  && \phantom{V_C(R)} = \frac{Z^2 e^2}{R}e^{-R/\lambda} 
                         f(R_p/\lambda) \;,
\label{Vc1}
\end{eqnarray}
where the dimensionless function $f(x)$ has been defined as follows:
\begin{equation}
 f(x)=\left[3\frac{x\cosh(x)-\sinh(x)}{x^{3}}\right]^{2} 
       \quad (x\equiv R_{p}/\lambda) \;.
\label{Littlef}
\end{equation}
Note that the function $f$ is independent of $R$. Indeed, it only
depends on the dimensionless ratio $R_{p}/\lambda$, namely, on the
interplay between the nuclear size and the screening length. In the
absence of screening, $f\!\equiv\!1$ (independent of nuclear
size) in accordance with Gauss' law. However, with screening $f$
becomes greater than one. The finite nuclear size places some of the
charges closer together than $R$; this increases the repulsion. Of
course, the finite size also places some of the charges farther
apart, thereby decreasing the repulsion. When these two effects are 
weighted by the screening factor $e^{-r/\lambda}$, the repulsion more 
than compensates for the ``attraction'' leading ultimately to
$f\!\ge\!1$. In the particular case of $R_{p}\!=\!5.45$~fm and 
$\lambda\!=\!10$~fm, one obtains $f(0.545)\!=\!1.061$ 
(about a 6\% increase).

\begin{figure}[ht]
\begin{center}
\includegraphics[width=2.75in,angle=270,clip=false]{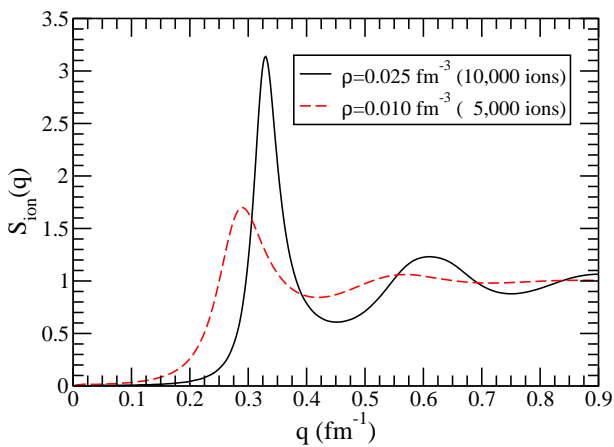}
\caption{(Color online) Ion static structure factor $S_{\rm ion}(q)$
as a function of the momentum transfer $q$. The black solid line is
from of a simulation with 10,000 ions corresponding to a density of
$\rho\!=\!0.025$~fm$^{-3}$. The red dashed line is from a simulation
with 5,000 ions corresponding to a density of $\rho\!=\!0.01$. See
text for details.}
\label{Fig18}
\end{center}
\end{figure}

The single-heavy-nucleus models consists of a gas of noninteracting
neutrons plus ions interacting via the screened Coulomb interaction
given in Eqs.~(\ref{Vc1}). Molecular dynamics simulations in the ion
coordinates are performed to compute its static structure factor
$S_{\rm ion}(q)$. The simulations used 5,000 to 10,000 ions and a time
step of 25 to 75 fm/c. The ion simulation can afford a larger time
step than the corresponding nucleon simulation because the heavier 
ions move slower. Further, the ion simulations require fewer particles 
to simulate the same physical volume because each ion ``contains''
several nucleons. The static structure factor for the ions
computed in this way (for densities of $\rho\!=\!0.01$~fm$^{-3}$ 
and $\rho\!=\!0.025$~fm$^{-3}$) is shown in Fig.~\ref{Fig18}.

Neutrino scattering from this system is described, in this 
single-heavy-nucleus model, by neutrino-nucleus elastic scattering 
within a framework that incorporates effects from both, the nuclear 
form factor and ion screening from the correlated nuclei. The cross 
section for elastic neutrino scattering from a single nucleus is 
proportional to the {\it square} of the weak charge of the nucleus 
times a suitable form factor to account for it finite size. For the
weak charge of the nucleus we simply use its neutron number $N$ as 
we continue to ignore the small weak charge of the proton, {\it i.e.,}
${\cal Q}_{\rm weak}\!=\!-N\!+\!Z(1\!-\!4\sin^{2}\theta_{\rm W})
\rightarrow -N$. Thus, the weak nuclear form factor reduces to that
of the neutron distribution. Further, to incorporate effects that
result from correlations among the ions, such as ion screening, the
cross section is multiplied by the ion static structure factor 
$S_{\rm ion}(q)$. Finally, one multiplies these terms by the fraction 
$X_h$ of heavy nuclei and divides by $N$ to obtain a static structure 
factor per neutron $S_{\rm model}(q)$ consistent with the
normalization of the earlier sections. That is,
\begin{equation}
  S_{\rm model}(q)=X_h N F(q)^2 S_{\rm ion}(q) \;.
\label{Smodel}
\end{equation}
Note that in addition to the coherent nuclear contribution there is a
small incoherent contribution from the neutron gas that has been
neglected. As defined above, this static structure factor can now be
directly compared to the one obtained in the full nucleon simulations.
{\it This prescription for $S_{\rm model}(q)$ corresponds to what is
presently used in most supernova simulations}. These simulations often
take $X_h$ and $N$ from the Lattimer-Swesty equation of
state~\cite{Lat92_NPA535} and $S_{\rm ion}(q)$ as computed in
Ref.~\cite{Hor97_PRD55}.

\begin{figure}[ht]
\begin{center}
\includegraphics[width=2.75in,angle=270,clip=false]{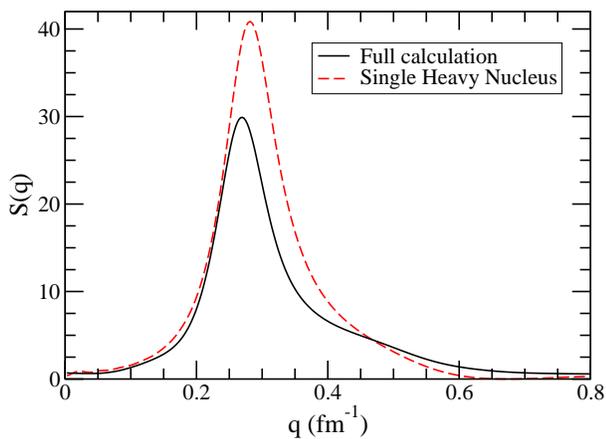}
\caption{(Color online) Neutron static structure factor $S(q)$ 
as a function of the momentum transfer $q$ at a density of 
$\rho\!=\!0.01$ fm$^{-3}$ for the full calculation (black solid 
line). Also shown (red dashed line) is the prediction from the 
ion static structure factor in Fig.~\ref{Fig18} including the 
square of the cluster form factor, as explained in the text.} 
\label{Fig19}
\end{center}
\end{figure}

In Fig.~\ref{Fig19} the model static structure factor 
$S_{\rm model}(q)$ is compared to the one from the full nucleon
calculation (see Sec.~\ref{results}) at a density of
$\rho\!=\!0.01$~fm$^{-3}$. The uniform form factor of Eq.~(\ref{F0})
is used with the sharp surface radius $R_n$ listed in 
Table~\ref{Table6}. For low to moderate momentum transfers the
agreement between the two approaches is excellent. This indicates
that---at this density and (low) momentum transfers---the system
is well described by a collection of nuclei of a single average mass. 
We expect that this good agreement will also hold at lower densities.  
However, there is a modest disagreement between $S_{\rm model}(q)$ and 
the complete $S(q)$ for $q\!>\!0.25$~fm$^{-1}$. This provides the
first indication of limitations within the single heavy nucleus 
approximation. The discrepancy could arise because the broad
distribution of cluster sizes displayed in Fig.~\ref{Fig13} is 
approximated by a single average cluster with a mass of $A=100$.  
Or it could be due to a breakdown in the factorization scheme.
That is, the cross section may no longer factor into a product
of a correlation function between ions ($S_{\rm ion}$) times the
weak response of a single ion ($N^{2}F(q)^{2}$).

\begin{figure}[ht]
\begin{center}
\includegraphics[width=2.75in,angle=270,clip=false]{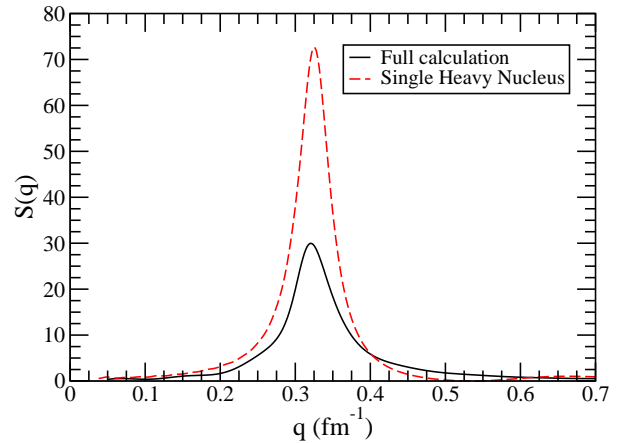}
\caption{(Color online) Neutron static structure factor $S(q)$ 
as a function of the momentum transfer $q$ at a density of 
$\rho\!=\!0.025$~fm$^{-3}$ (black solid line). Also shown (red 
dashed line) is the prediction from the ion static structure 
factor in Fig.~\ref{Fig18} including the square of the cluster 
form factor, as explained in the text.} 
\label{Fig20}
\end{center}
\end{figure}

A similar comparison is done in Fig.~\ref{Fig20} but now at the higher
density of $\rho\!=\!0.025$~fm$^{-3}$. Now the disagreement between
$S_{\rm model}(q)$ and $S(q)$ is more severe. This indicates that
errors in the single nucleus approximation will grow rapidly with
density. Moreover, the single nucleus approximation {\it overpredicts}
the neutrino opacity relative to the complete calculations.

\section{Conclusions}
\label{conclusions}

Nonuniform neutron-rich matter was studied via semiclassical
simulations with an interaction that reproduces the saturation density
and binding energy of nuclear matter and incorporates the long-range
Coulomb repulsion between protons. Simulations with a large number of
nucleons (40,000 to 100,000) enable the reliable determination of the
two-neutron correlation function and its Fourier transform---{\it the
static structure factor}---even for low momentum transfers. The static
structure factor $S(q)$ describes coherent neutrino scattering that is
expected to dominate the neutrino opacity. At low momentum transfer
$q$ the static structure factor is small because of ion screening;
correlations between different clusters screen the weak charge. At
intermediate momentum transfers a large peak is developed in $S(q)$
corresponding to coherent scattering from all of the neutrons in a
cluster. This peak moves to higher $q$ and decreases in amplitude as
the density of the system increases.

In principle the neutrino opacity could be greatly increased by large
density fluctuations. A simple first-order phase transition has a
two-phase coexistence region where the pressure is independent of
density. Large density fluctuations in this region imply a very large
value of the static structure factor at very small momentum
transfers. Indeed, $S(q\!=\!0)$ is directly proportional to the
density fluctuations in the system. Moreover, density fluctuations are
also proportional to the isothermal compressibility.  For consistency,
the static structure factor $S(q\!=\!0)$ was computed in these two
equivalent yet independent ways, namely, as the Fourier transform of
the two-neutron correlation function and as the derivative of the
pressure with respect to the baryon density. While we find good
agreement between these two schemes, no evidence is found in favor of
a large enhancement in $S(q\!=\!0)$. We conclude that the system does
not undergo a simple, single component first-order phase transition,
so no large increase in the neutrino opacity was found.

To compare our simulation results to more conventional approaches of
wide use in supernova calculations a cluster model was introduced. A
{\it minimal spanning tree clustering algorithm} was used to determine
the composition of the various clusters (``nuclei'') in the
simulations (see for example Ref.~\cite{Pap98_Dov}).  To make contact
with some of these conventional approaches, such as the {\it single
heavy nucleus approximation}, the neutrino opacity was computed in a
system modeled as a gas of free neutrons and a representative ({\it
i.e.,} average) single-species heavy nucleus. The neutrino opacity for
such a system is dominated by elastic scattering from the heavy
nucleus. The contribution from the single nucleus to the neutrino
response is proportional to the square of its weak charge (assumed to
be carried exclusively by the neutrons) and its elastic neutron
form-factor, that accounts for its finite size. Further, Coulomb
correlations among the different nuclei was incorporated through an
{\it ion static structure factor} to account for ion screening.
Fairly good agreement is found between the single heavy nucleus
approximation and our complete simulations at low density and
especially at small momentum transfers. However, starting at a density
of approximately $10^{13}$~g/cm$^3$, we find a large disagreement
between the two approaches that grows rapidly with increasing density.
In particular, our complete simulations yield neutrino opacities that
are smaller than those in the single heavy nucleus approximation. Note
that our full simulations yield accurate results even at the (high)
densities where the single heavy nucleus approximation becomes
invalid. We reiterate that the single heavy nucleus approximation is
what is presently employed in most supernova simulations.

Future work could include calculating the dynamical response of the
system to study the transfer of energy between neutrinos and matter.
Note that a great virtue of molecular dynamics approaches combined
with special purpose computers (such as what has been done here) is
that dynamical information for systems with large number of particles
may be readily obtained from time-dependent correlations. Particularly
interesting is the low-energy part of the response which should be
dominated by the so-called {\it Pygmy resonances}. These oscillations
of the neutron skin of neutron-rich nuclei against the symmetric core
should be efficiently excited by low-energy neutrinos. Another
promising area for future research is the spin response of the
system. A first step could involve including spin dependent forces 
in our model. The spin response is interesting because nucleons have 
large spin dependent couplings to neutrinos.

\appendix*
\section{MDGRAPE}
To do the simulations we used a special purpose computer called the
MDGRAPE-2. The MDGRAPE-2 is a single board which plugs into the PCI bus
of a general purpose computer, and is designed for extremely fast
calculation of forces and potentials in molecular dynamics simulations
\cite{Nar99_MS21}. It is the third generation of such hardware, which
evolved from the work of J. Makino et al at the University of Tokyo on
similar hardware called the GRAPE (for GRAvity PipE), for doing
gravitational N-body problems \cite{Mak00_SC2000}. In our case, we have
two boards plugged into the PCI bus of one of the Power3+ nodes of
Indiana University's IBM SP supercomputer. Each board is rated at a peak
speed of 64 GigaFLOPS (floating point operations per second). The
MDGRAPE-2 can compute any central potential of the form
\begin{equation}
   V(i,j) = b_{ij} f(a_{ij}(r_{ij}^2+\epsilon_{ij}^2))
                                              \label{eq:MDGRAPE-pot}
\end{equation}
or the corresponding central force, which is of the same form, except
multiplied by $\rvec$. All three terms in (2) are of this form. In our
case $\epsilon_{ij}=0$, and $b_{ij}$ and $a_{ij}$ are either scalars,
or $2\times 2$ matrices, corresponding to the two particle types proton
and neutron. The boards are accessed via the M2 library, which the user
links into his code. The library is very easy to use, and handles
distribution of work between the two MDGRAPE-2 boards without user
intervention. The user defines $f(x)$ by a function table of 1024 points,
which the MDGRAPE-2 interpolates via fourth degree polynomial interpolation.
One can thus reproduce most physically realistic functions very accurately.
Software is provided for constructing function tables, which are stored
in files and loaded in during runtime.

At each MD time step, one calls M2 subroutines to load in the function
table and the scale factors or matrices $a_{ij}$ and $b_{ij}$. One then
calls a subroutine to load in the source particle coordinates, and
subroutines to load integer arrays of particle types (0 for neutron, 1
for proton) for both source and target particles. Then one calls a force
calculation routine, passing it the array of target particle coordinates.
In our case the source and target particles are the same, but they do not
have to be. One input parameter to the force calculation specifies that
periodic boundary conditions should be used. The MDGRAPE-2 has built in
hardware for taking periodic b.c. into account. The output is an array
containing the total force on each target particle. A similar call can
be made to compute the total potential energy of each particle.

We must go through these steps three times, once for each term in (2).
Still, the MDGRAPE-2 is much faster than serial Fortran code. In ordinary
Fortran, this whole calculation would be done in a pair of nested DO
loops, and thus would take of order $O(A^2)$ time. For our simulations
with $A=40,000$ this would be prohibitive even for today's fast CPUs.
But the two MDGRAPE-2 boards together can do the force calculation about
90 times faster than a single Power3+ processor, so that a simulation of
100,000 MD time steps that would take over two years using a serial
program can be done in less than nine days. Benchmark tests show this
speedup holds out to at least $A=160,000$. Each MDGRAPE-2 board has
enough memory to hold a half million particles, so we have not yet
reached our maximum capability.

We calculated the neutron-neutron
correlation function $g(r)$ (see below) using ordinary Fortran code, as
it was not clear how to perform this calculation with the MDGRAPE-2.
Although $g(r)$ is also an $O(A^2)$ calculation, it is done infrequently,
and does not severely impact performance.

\begin{acknowledgments}
We acknowledge useful discussions with Sanjay Reddy.
M.A.P.G. acknowledges partial support from Indiana University and
University of Oviedo.  J.P. thanks David Banks and the staff at the
FSU Visualization Laboratory for their help.   We thank Brad Futch for preparing Figs. 3,7,11.  This work was supported in part by DOE grants DE-FG02-87ER40365 and DE-FG05-92ER40750, and by Shared University Research grants from IBM, Inc. to Indiana University.
\end{acknowledgments}

\vfill\eject

\bibliography{ReferencesJP}

\end{document}